\documentclass{article}
\usepackage{spconf,amsmath,amssymb,graphicx}
\usepackage{color}
\usepackage{balance}
\usepackage{multirow}

\newcommand{\diag}{\mathop{\mathrm{diag}}\nolimits}
\usepackage[]{algorithm2e}
\usepackage{wasysym}
\usepackage{bm}
\usepackage{url,cite}
\usepackage{enumitem} % tune margin for \begin{itemize}
\usepackage{pifont}% http://ctan.org/pkg/pifont

\def\tthline{\noalign{\hrule height 1.4pt}}
\def\thline{\noalign{\hrule height 1.0pt}}

\title{Bootstrapping single-channel source separation via\\unsupervised spatial clustering on stereo mixtures}
\name{Prem Seetharaman$^{1}$, Gordon Wichern$^2$, Jonathan Le Roux$^2$, Bryan Pardo$^1$}
\address{
$^1$Northwestern University, Evanston, IL, USA\\
$^2$Mitsubishi Electric Research Laboratories (MERL), Cambridge, MA, USA\\
}

\begin{document}
\ninept
\maketitle
\setlength{\abovedisplayskip}{4pt}
\setlength{\belowdisplayskip}{4pt}
\begin{abstract}
Separating an audio scene into isolated sources is a fundamental problem in computer audition, analogous to image segmentation in visual scene analysis. Source separation systems based on deep learning are currently the most successful approaches for solving the underdetermined separation problem, where there are more sources than channels. Traditionally, such systems are trained on sound mixtures where the ground truth decomposition is already known. Since most real-world recordings do not have such a decomposition available, this limits the range of mixtures one can train on, and the range of mixtures the learned models may successfully separate. In this work, we use a simple blind spatial source separation algorithm to generate estimated decompositions of stereo mixtures. These estimates, together with a weighting scheme in the time-frequency domain, based on confidence in the separation quality, are used to train a deep learning model that can be used for single-channel separation, where no source direction information is available. This demonstrates how a simple cue such as the direction of origin of source can be used to bootstrap a model for source separation that can be used in situations where that cue is not available.
\end{abstract}
\vspace{-.2cm}
\section{Introduction}
\vspace{-.2cm}
Separating an audio scene into isolated sources is a fundamental problem in computer audition, analogous to image segmentation in visual
scene analysis. Robust source separation would improve many technologies, including hearing aids, speech recognition in complex auditory environments, and biodiversity monitoring (e.g., birdsong identification).

Source separation systems based on deep learning are currently the  most successful methods for separating recordings containing multiple concurrent sounds in underdetermined conditions, that is, where there are fewer channels than sources \cite{stoter2018sisec}.  Traditionally, deep learning systems are trained on many mixtures (e.g., tens of thousands) for which the ground truth decompositions are already known. Since most real-world recordings have no such decomposition available, developers train systems on artificial mixtures created from isolated individual recordings. Although there are large databases of isolated speech, it is impractical to find or build large databases of isolated recordings for every arbitrary sound. This fundamentally limits the range of 
sounds that deep models can learn to separate. 

The traditional learning procedure for these source separation models is in contrast to how humans learn to segregate audio scenes \cite{bregman1994auditory}: sources are rarely presented in isolation and almost never in ``mixture/reference'' pairs. One can argue that the brain is able to learn to separate sounds without having access to large datasets of isolated sounds. There is experimental evidence that the brain uses fundamental cues (e.g., direction of origin of a sound) that are independent of the characteristics of any particular sound source to perform an initial segmentation of the audio scene \cite{mcdermott2011empirical}. The brain could use such cues to separate at least some scenes to some extent, and use that information to train itself to separate more difficult scenes.  

In many stereo recordings (both natural and artificial), sources are spatialized such that the primary signal energy from one source comes from a different direction than that of another source. In a stereo (a.k.a. two-channel) recording, the direction of origin of a source is typically manifested as a phase and amplitude difference between the two channels. Source separation approaches such as DUET \cite{rickard2007duet} and PROJET \cite{fitzgerald2016projet} have exploited such differences to perform separation without relying on training data. 

In this work, we explore using spatial source separation on stereo mixtures to generate initial decompositions of audio mixtures. The decompositions vary greatly in quality from mixture to mixture. We derive a confidence measure in the decompositions, based on the clustering of features of the stereo mixture. These decompositions are weighted by confidence and used to train a deep-learning source separation model, here based on deep clustering \cite{hershey2016deep}. Once trained, the model can be applied to separate single-channel mixtures, where no source direction information is available. The idea is to use simple, low-level processing to separate sources using spatial information in easy conditions (e.g., where the sources are well separated spatially and reverberation is limited) and then use that knowledge to bootstrap a source separation model for difficult conditions.

Several recent efforts have attempted to learn to perform tasks such as representation learning or source separation in one modality by using another modality to perform cross-modality self-supervision. In the case of audio, these works learn to localize or separate sounds by using vision as the extra modality \cite{aytar2016soundnet,zhao2018sound,owens2018eccv,arandjelovic2017objects,gao2018learning}. In contrast, our work explores the use of stereo audio to supervise single-channel audio source separation, instead of crossing modalities.

\begin{figure}
    \centering \vspace{.2cm}
    \includegraphics[width=.99\linewidth]{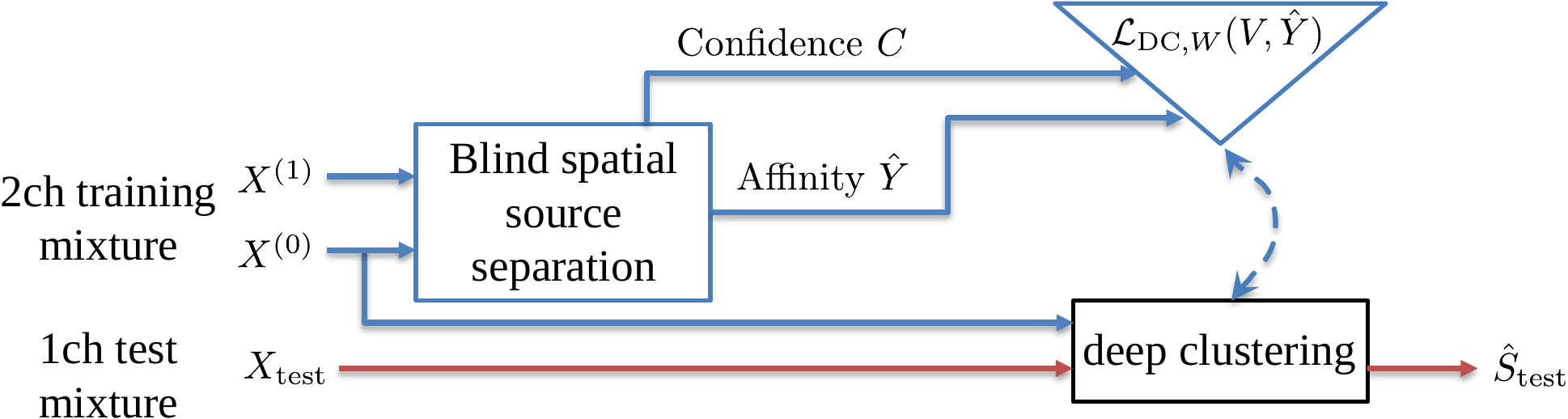}
    \vspace{.05cm}
    \caption{Illustration of the proposed bootstrapping of single-channel separation using blind spatial separation.}\vspace{-.35cm}
    \label{fig:system}
\end{figure}

This work can also be considered as an instance of deep learning in the presence of noisy labels, which has previously been explored for images \cite{reed2014training, sukhbaatar2014learning}. The estimation of confidence measures of source separation estimates was explored in \cite{manilow2017predicting} and learning to separate sources using features derived from cues was explored in \cite{bach2006learning}. In contrast to \cite{manilow2017predicting}, we derive a confidence measure based on the clustering space and do so without requiring any training. In contrast to \cite{bach2006learning}, we treat the output of the spatial cue as ``pseudo ground truth'', rather than using it as an input feature and mapping it to the actual ground truth. The system is illustrated in Fig.~\ref{fig:system}.

\section{Proposed Method}
\subsection{Spatial source separation}
To generate the initial segmentations used for training the model, we use a simple blind source separation method that clusters time-frequency (T-F) bins based on low-level features present in stereo mixtures. This method belongs to a well studied family of spatial source separation algorithms \cite{vincent2007first} such as DUET \cite{rickard2007duet} and GMM spatial clustering \cite{kim2011gaussian}. The  assumption is that T-F bins with similar spatial features likely come from the same direction, and that sounds coming from the same direction belong to the same source. If the sources are coming from distinct spatial locations, one will observe significant inter-channel difference, giving a good clustering and separation result.
The key idea is to exploit differences between the two channels to decide which T-F bins go with which source. We first transform the input stereo audio to a stereo complex spectrogram $X^{(c)}_{t, f}$ where $c$ is the channel, $t$ the time index, and $f$ the frequency index. We then extract the interchannel phase difference (IPD) $\theta$ and the log magnitude spectrogram $X^{\log}$:
\begin{align}
    \theta_{t, f} &= \angle\Big(X^{(0)}_{t, f}\overline{X^{(1)}_{t, f}}\Big), \\
    X^{\log}_{t,f} &= 20\log_{10}\big({|X^{(0)}_{t, f}|}\big).
\end{align}

We use the cosine and sine of the IPD, $\text{cosIPD}=\cos{\theta_{t, f}}, \text{sinIPD}=\sin{\theta_{t, f}}$ to form a two dimensional feature space. As these features are correlated, we project them down to a single dimension, $\phi_{t, f}$, using principal component analysis (PCA).
We cluster the feature space using a Gaussian mixture model (GMM) with a full covariance matrix that is fit using the expectation-maximization (EM) algorithm. We use such a clustering approach because it lets us derive a confidence measure for the assignment of T-F bins to sources. 

To bias the clustering towards bins with significant energy, we only fit the GMM to bins such that $X^{\log}_{t,f} > \tau$, where $\tau$ is a manually set threshold (set to $-10$ in this work). The number of components $N$ in the GMM is set ahead of time ($N=2$ in this work). The GMM posterior assignments are used as masks on the complex spectrogram, one for each Gaussian component $z_j$:
\begin{equation}
    \gamma^{(j)}_{t, f} = \frac{P(\phi_{t, f}| z_{j})P(z_j)}{P(\phi_{t, f})}.
\end{equation}

We use the spatial information contained in stereo recordings to estimate a (pseudo) label matrix $\hat{Y}$. We do this by comparing the masks produced by the GMM. The mask with the highest value (i.e. highest posterior probability) for a T-F point determines the label of that point.
Given T-F bins $i$ and $k$, the value for $\hat{Y}_{i,k}$ is binary: $1$ if they belong to the same source, and $0$ if they belong to different sources. This is done the same way as in the original deep clustering work \cite{hershey2016deep}, with binary masks, except here the estimated sources from the spatial model are used as pseudo ground truth.

\subsection{Confidence measure}
 \vspace{-.2cm}

Compared to the ground truth label matrix $Y$, assignments in $\hat{Y}$ may be incorrect. As we do not have access to ground truth, we derive a confidence measure from the Gaussian mixture model fit to the spatial features. We measure three aspects of the clustering to compute an overall confidence: cluster size equality, clustering fit, and posterior assignments.

\textbf{Cluster size equality:} The $N$ clusters should contain a roughly equal fraction of the total number $TF$ of T-F bins, where $T$ and $F$ are the number of time frames and frequencies, to mitigate mode collapse (all points being assigned to one cluster). This scalar measure is defined as: 
    \begin{equation}
        C_{\text{cl}} = \sum_{j=1}^N \Big(\frac{1}{N} - \Big|\frac{1}{N} - f_j\Big|\Big).
        \label{eq:cluster}
    \end{equation}
    where $f_j$ is the fraction of T-F bins hard-assigned to cluster $j$.
    
    %cluster_sizes = np.sum(labels[fit_weights], axis=0)
    %cluster_sizes /= (np.sum(cluster_sizes) + 1e-6)
    %cluster_size_weight = ((source_share - np.abs(cluster_sizes - source_share)) * 2)[0] + 1e-3
    
\textbf{Clustering fit:} To measure how well separated the clusters are in the spatial feature space, we compute the Jensen-Shannon Divergence (JSD) \cite{lin1991divergence} between a GMM $P$ with one component and a GMM $Q$ with $N$ components, both fit to that space.
With KL denoting the KL-divergence, this scalar measure is defined as:
    \begin{equation}
        C_{\text{JSD}} = \text{JSD}(P \parallel Q)= \frac{1}{2}\text{KL}(P \parallel \frac{P+Q}{2})+\frac{1}{2}\text{KL}(Q \parallel \frac{P+Q}{2})
        \label{eq:jsd}
    \end{equation}
    %formula goes here
If the spatial features cluster into fewer than $N$ distinct sources, the overlap between the $N$ component GMM and the single component GMM will increase. The JSD measures this and returns a number between $0$ (completely overlapping distributions) and $1$ (distinct distributions).  Note that Eq.~\eqref{eq:jsd} has no closed-form solution for arbitrary mixture models \cite{hershey2007approximating, durrieu2012lower}. Instead of computing \eqref{eq:jsd} directly, we approximate it using the Monte Carlo method.
Finally, JSD was chosen over other information criteria, such as Bayesian and Akaike, as it does not penalize for the number of parameters and also actually computes the overlap between the distributions rather than the difference in log likelihood.

\textbf{Posteriors:} We use the posteriors $\gamma$ to measure how confident the GMM is for each T-F bin, unlike the previous two global measures. Points with unsure posteriors (assignment shared roughly equally by all components) are down-weighted.
This measure is defined as:
    \begin{equation}
        C_{\text{post}}(t, f) = 2{\Big|\max_{j \in \{1,\dots,N\}}\gamma^{(j)}_{t, f}} - \frac{1}{2}\Big|
        \label{eq:post}
    \end{equation}
    %posterior_scores = (2*np.abs(np.max(assignments, axis=-1) - .5))

\begin{figure}
    \centering
    \includegraphics[width=\linewidth]{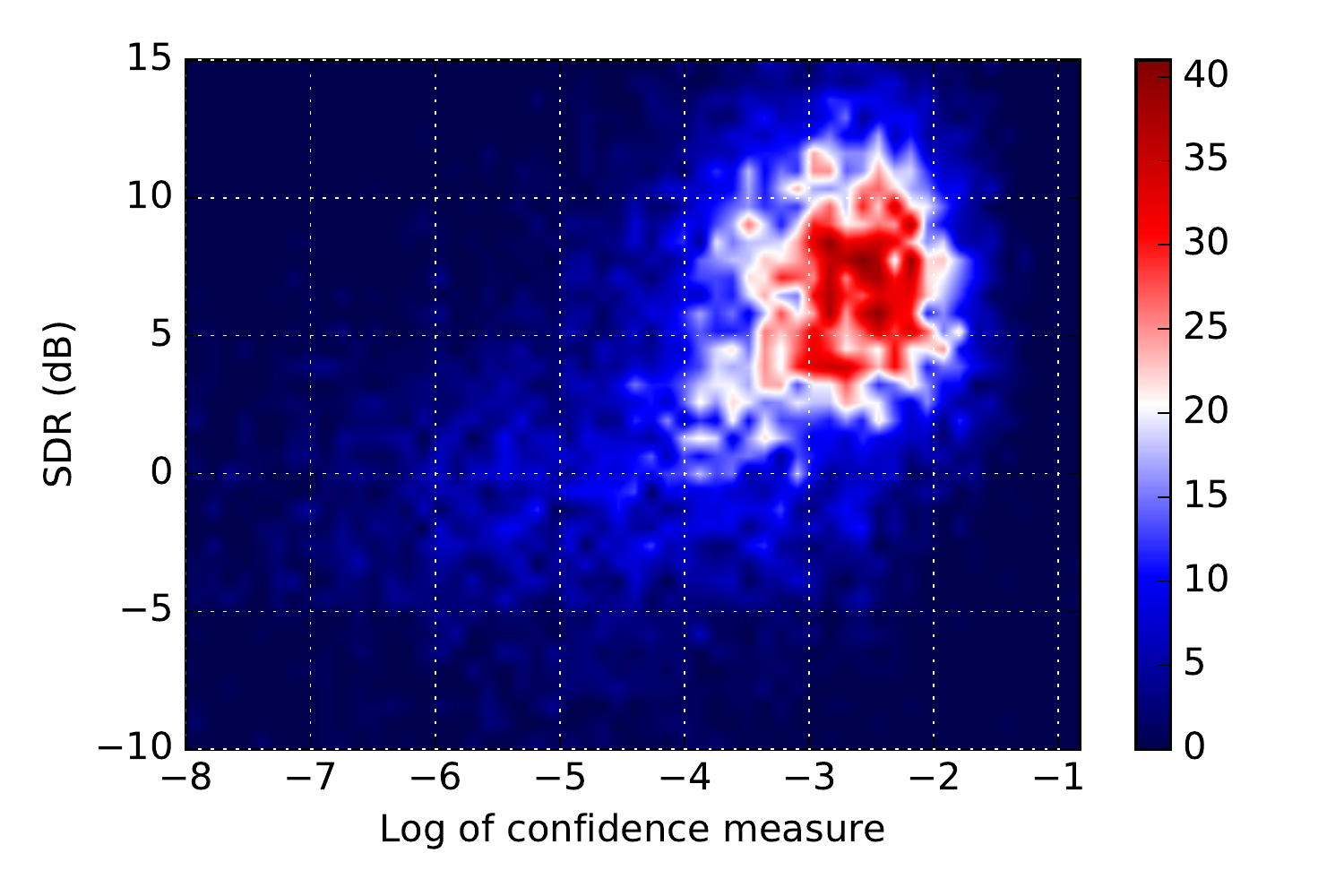}
    \vspace{-.9cm}
    \caption{Relationship between log of confidence measure and SDR. }\vspace{-.5cm}
    \label{fig:confidence_measure}
\end{figure}

Equations \eqref{eq:cluster}, \eqref{eq:jsd}, and \eqref{eq:post} all produce numbers in $[0,1]$. To compute the overall confidence measure $C$, we simply take the product of the three measures and raise to a power $\alpha$:
\begin{equation}
    C_{t, f}(\alpha) = [C_{\text{cl}}C_{\text{JSD}}C_{\text{post}}(t, f)]^\alpha. \label{eq:confidence}
\end{equation}
This confidence measure weights every time-frequency point in the representation. It ranges between $0$ (low confidence) and $1$ (high confidence). The exponent $\alpha$ is a tunable parameter that can be used to emphasize 
or de-emphasize 
high confidence examples. We test $\alpha = 0.5, 1, 2$. In Fig.~\ref{fig:confidence_measure}, we show the relationship between the confidence measure for a mixture and the source-to-distortion ratio (SDR) for the separation for the validation mixtures in our dataset (Section \ref{sec:dataset}). We plot the log to visualize the lower end of the distribution. The correlation between the confidence measure (not its log) and SDR has an r-value of $0.36$ with a p-value $\ll0.001$.

\vspace{-.2cm}

\subsection{Training the single-channel model}
\vspace{-.2cm}
\label{subsection:training}
The model we use for source separation is based on deep clustering (DC) \cite{hershey2016deep,isik2016single}. We selected deep clustering because it is a highly successful approach that has inspired multiple successful variants \cite{chen2017deep, luo2017deep, kumar2018music, settle2018end, wang2018alternative, Wang2018Interspeech09}. Further,  its separation framework is somewhat connected to our primitive spatial separation as it is based on clustering as well, but in a learned embedding space, and its objective function has been shown to be amenable to the introduction of weights.
In DC, a neural network is trained to map each T-F bin in a spectrogram to a higher-dimensional embedding where bins that belong to the same source are near each other and bins that belong to different sources are far from each other. Once trained, the network is used to embed a new spectrogram representing an auditory scene. Sources are then recovered by clustering in the embedding space. 

Here, we train the DC network using a label matrix obtained by treating the estimates from the spatial segmentation as pseudo ground truth. Because this pseudo ground truth may not be reliable, we use a version of the  original DC objective \cite{hershey2016deep} that was modified to include a weight $w_i$ for each T-F bin $i=(t,f)$ \cite{wang2018alternative}: 
\begin{align}
    \mathcal{L}_{\text{DC},W}(V,Y) &= \|W^{1/2} ( VV^T - YY^T) W^{1/2}\|_{F}^2 \label{eq:sqrtW}\\
    &= \sum_{i, j} w_i w_j[\langle v_i, v_j \rangle - \langle y_i, y_j \rangle]^2,
    \label{eq:dpcl}
\end{align}
where $V=(v_i)$ denotes an embedding matrix, $Y=(Y_i)$ a label matrix, and $W=\diag(w)$ a diagonal matrix with the weights on the diagonal.
In \cite{wang2018alternative}, weights were introduced to make the network focus on louder T-F points in the spectrogram, since these points have a bigger effect on perceived separation quality, and assignments of silent regions are rather arbitrary and should thus not have a large impact when learning the embeddings. 
Here, we use the weighted version of the DC loss function, but our weights instead incorporate both the confidence measure $C$ and the magnitude weighting. Specifically, $w_{t, f}(\alpha) = C_{t, f}(\alpha)[|X_{t, f}|/\sum_{t,f} |X_{t,f}|]$. This objective function makes the network focus on learning embeddings for T-F points that are both classified by the spatial classifier with high confidence and also have significant energy. Because spatial information is only used in the objective function, once the network is trained, we can use it to process single-channel mixtures (where the spatial source separation algorithm cannot be used) and cluster the embeddings using K-means clustering to recover the sources.

\vspace{-.2cm}
\section{Experiments}
\vspace{-.2cm}
We investigate whether single-channel source separation can be bootstrapped from noisy estimates produced by a stereo separation algorithm, and whether weighting the estimates using confidence improves performance of the bootstrapped model. We also explore ensembles of the spatial algorithm and the bootstrapped models.

\vspace{-.2cm}
\subsection{Network architecture}
Our network is similar to those in the original deep clustering works \cite{hershey2016deep, isik2016single}, consisting of a four layer BLSTM stack with 300 units in each direction followed by a dense layer.  The dense layer uses a $\mathrm{tanh}$ non-linearity and outputs a $15$ dimensional embedding for each T-F bin. The network has $8.7$M parameters. %$8691797$. 
This network architecture was used for all models, with the same random initialization.

\vspace{-.2cm}
\subsection{Dataset and training procedure}
\label{sec:dataset}
Our training, validation and test data are from the publicly available\footnote{\url{http://www.merl.com/demos/deep-clustering}} spatialized version of the Wall Street Journal mix dataset with two speakers (wsj0-2mix) \cite{hershey2016deep, wang2018multi}. This dataset is created by randomly mixing the speakers at random locations in synthetic rooms in reverberant and anechoic conditions. We use the anechoic version of this dataset in this work, where the speakers are panned at random (sometimes overlapping) angles. There are 20000, 5000, and 3000 two-speaker 8-channel mixtures for training, validation, and testing. Our spatial algorithm only operates on two of the first 4 channels, randomly selected to create a stereo mixture at training time. Our deep clustering model is trained on the single-channel mixture corresponding to the first channel of the stereo mixture.

We consider two possible outputs for training our model. The first is the ground truth decomposition, which are available because our dataset contains separated sources. The second is the estimated decomposition provided by the spatial source separation algorithm. This algorithm is based on inter-channel phase difference, clustering, and time-frequency masking and does not achieve great separation quality. The motivation of this work is to see if it is possible to learn how to perform source separation from a biologically inspired source separation algorithm that produces noisy estimates in concert with a confidence weighting scheme. Due to the poor performance of the spatial algorithm on reverberant data ($1.1$ dB SDR), we restrict our analysis to the anechoic case. We hypothesize that a better blind spatial source separation algorithm that can handle reverberant cases would allow for even more successful bootstrapping of a model.

The audio mixtures have a sampling rate of 8~kHz, spectrogram window size of 32 ms
and hop size of 8 ms.
The input to the network is a sequence of log magnitude spectrogram features, with sequences of at most 400 frames used for training.  The networks are trained for $100$ epochs with a batch size of $40$ and optimized using Adam with an initial learning rate of 1e-3. The learning rate is decayed by half if the validation loss does not go down for $5$ consecutive epochs. 

\begin{figure}
    \centering
    \includegraphics[width=\linewidth]{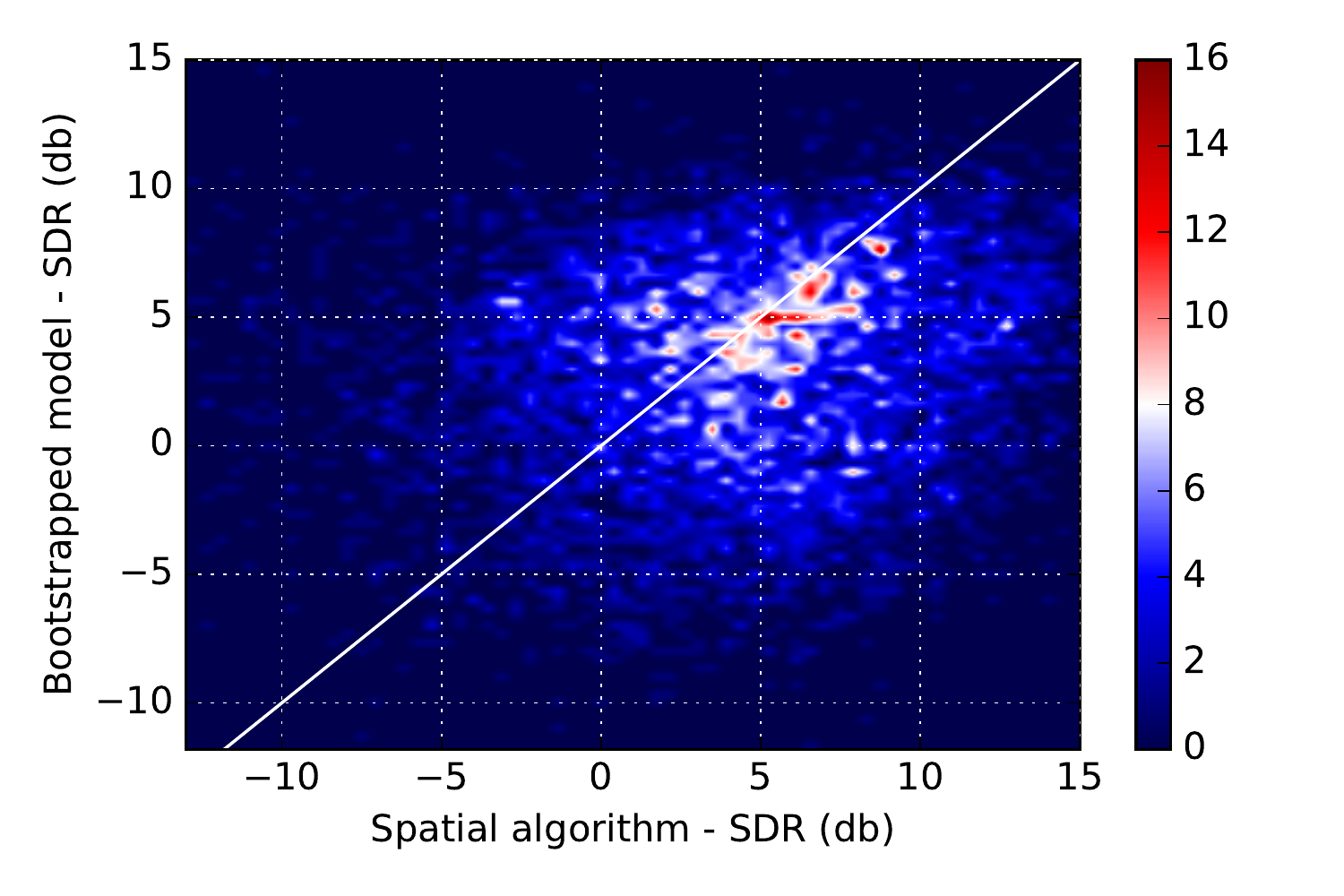}\vspace{-.5cm}
    \caption{Relationship between performance of the spatial algorithm and the bootstrapped model with the confidence weighting scheme ($\alpha = 1$) on test mixtures, with the line $y=x$ plotted in white.}\vspace{-.45cm}
    \label{fig:sdr_comparison}
\end{figure}

%\vspace{-.2cm}
\subsection{The source separation approaches we compared }
\vspace{-.05cm}
We trained a set of deep clustering (DC) models that all share the same architecture and initialization weights, but were trained either on ground truth separated signals (providing an upper bound on performance) or on source separation results produced by the spatial separation model. Unless otherwise noted, all models were trained until convergence on all 20k training examples. Models trained on the spatial model output were either provided raw training examples (with only magnitude weighting via $w(0)$) or examples weighted using our confidence measure via weights $w(\alpha)$ with $\alpha>0$. 

We tested each of the trained models on single channel mixtures: the first channel in each of the $3000$ test mixtures. We also evaluated performance of the spatial separation algorithm. For the spatial model, we took the first channel and a random other channel from the other $3$ channels to produce a stereo mixture. We make a mask using the spatial algorithm and applied it to the first channel. 

Since the confidence measure relies only on the unsupervised spatial separation algorithm, it does not require ground truth to be computed. This lets us explore using the best bootstrapped model (DC on estimates, weighted, $\alpha = 1$) in concert with the spatial source separation on stereo mixtures, using the mean confidence measure to mediate between the two approaches.   
If confidence in the spatial model's output is low, we discard it and use the DC model. If confidence is high, we discard the DC model and use only the spatial model. We set the switching point at the bottom quartile of all the mean confidence measures across all validation mixtures. We compared this approach (confidence) to one where the true performance of both approaches is known and the best output is always selected (oracle) and one where the approach is selected randomly.

\vspace{-.2cm}
\subsection{Results}
 
{\setlength{\tabcolsep}{3pt}
\begin{table}[t]
\footnotesize
\centering
\caption{SI-SDR (dB) for each approach. DC: Deep clustering. 
}
\label{tab:results}
\begin{tabular}{cl|cc|ccc}
\tthline
&Approach              & Quality & Quantity & SDR & SIR  & SAR  \\ \tthline
\multirow{6}{*}{1ch}&DC w/ ground truth (20k)  &1.000&1.000    & $9.2$   & $22.4$  & $9.5$   \\ 
&DC w/ ground truth (1k)   &1.000&0.050    & $2.5$   & $11.5$  & $3.3$   \\
\cline{2-7}
&DC w/ estimates, $\alpha=0$ &0.303&1.000& $1.8$   & $11.9$  & $2.8$   \\
&DC w/ estimates, $\alpha=0.5$ &0.362&0.205& $2.8$   & $13.5$  & $3.6$   \\ 
&DC w/ estimates, $\alpha=1$  &0.387&0.054 & ${\bf2.9}$   & ${\bf 13.5}$  & ${\bf 3.7}$   \\
&DC w/ estimates, $\alpha=2$ &0.400&0.005 & $1.8$   & $11.5$  & $2.7$   \\\thline
\multirow{4}{*}{2ch}&Spatial algorithm    &-&-    & $4.3$   & $17.3$  & $5.6$   \\ 
\cline{2-7}
&Ensemble (oracle, $\alpha=1$)    &-&-    & $6.2$   & $19.6$  & $7.0$   \\ 
&Ensemble (random, $\alpha=1$) &-&-    & $3.6$   & $15.4$  & $4.6$   \\ 
&Ensemble (conf., $\alpha=1$)  &-&-   & ${\bf 5.0}$   & ${\bf 18.3}$  & ${\bf 5.9}$   \\ \tthline
\end{tabular}\vspace{-.3cm}
\end{table}}

Table \ref{tab:results} shows the performance of each algorithm in terms of scale-invariant SDR (SI-SDR) \cite{LeRouxSDR}. % (Table \ref{tab:results}).
We first observe that deep clustering trained on the ground truth far outstrips the other approaches, indicating that there is still work to be done to bootstrap high quality source separation models. We also see that the confidence weights have a  significant impact on performance of the bootstrapped model, raising it by $1.1$ db SDR. This indicates that confidence weighting is important for learning from estimates. This could be because it increases the signal to noise ratio in the training data. The exponent $\alpha$ in Eq.~\eqref{eq:confidence} controls the balance between quality and quantity of training data effectively seen by the model:
with $\alpha=0$, all data is considered as pseudo ground truth, regardless of quality, while higher values of $\alpha$ de-emphasize low-confidence examples, improving average quality at the expense of the effective total amount of training data. While $\alpha = 0.5$ and $\alpha = 1$ show good performance,  both the low value $\alpha = 0$ and high value $\alpha = 2$ lead to significantly decreased performance, indicating a trade-off between quantity and average quality of the examples. 

We estimate the proportion of the training data used by the bootstrapped model ($\alpha=1$) versus the ground truth model, by comparing the sum of confidence weights $w_i(\alpha)$ across the entire dataset. This quantity measure, shown in Table \ref{tab:results}, indicates the bootstrapped model effectively sees about $5$\% of the training data seen by the ground truth model. When trained on $5$\% of the training data (1k examples), the SDR of the ground truth model decreases from $9.2$ dB to $2.5$ dB, on par with the bootstrapped model with $\alpha\!=\!1$. As alpha increases, the amount of effective training data decreases.

To quantify the quality of the labels seen by the model, we use $1-d_{\chi^2}(w(\alpha)\odot Y, w(\alpha)\odot \hat{Y})$ where $d_{\chi^2}$ is the chi-squared distance between partitions \cite{Hubert1985,bach2006learning,meilua2012local,wang2018alternative}, applied between the ground truth labels $Y$ and the estimated labels $\hat{Y}$ produced by the spatial algorithm, where the label matrices are weighted by $\sqrt{w_i(\alpha)}$ %(for $\alpha=0,0.5,1,2$) 
at each T-F point (similarly to Eq.~\eqref{eq:sqrtW}). We compute the weighted average across the entire training dataset, with each example weighted by the sum of confidence weights $\sum_i w_i(\alpha)$ over that example, as the quality shown in Table \ref{tab:results}. As expected, quality increases with $\alpha$. 

The spatial algorithm outperforms the bootstrapped model, although the comparison is not fair because the spatial algorithm has stereo input while the bootstrapped model has only single-channel input. In single-channel cases or cases with little spatial separation, the spatial model cannot be used at all. Figure \ref{fig:sdr_comparison} shows the relationship between the SDRs for both approaches. There are many cases where one approach is better than the other, indicating an ensemble approach may be fruitful in the stereo setting. This is akin to the human auditory system, which mediates between primitives (cues) and schema (learned models) to successfully parse the auditory scene \cite{bregman1994auditory}. Table \ref{tab:results} shows an ensemble method (relying on stereo cues when the confidence measure is high and switching to the bootstrapped model if low) out-performs either approach in isolation. This indicates that in difficult cases where the spatial algorithm fails, the bootstrapped model is more successful, on average. An ensemble that picks randomly between the spatial algorithm and the bootstrapped model under-performs the spatial algorithm by itself, indicating the usefulness of the confidence measure to control selection. An oracle ensemble improves on the confidence ensemble by $1.1$ db, suggesting room for improvement in the confidence measure.

\section{Conclusion}
We have presented a biologically inspired method for bootstrapping a single-channel deep network for source separation. The model is trained on noisy separation estimates produced by a spatial audio source separation algorithm applied to stereo mixtures. The trained model can separate sources in single-channel mixtures, where the cue needed by the method that trained the model is not present.  We constructed a confidence measure in the output of the spatial algorithm. A similar confidence measure can be defined for any clustering-based separation algorithm. We use this measure to reduce the impact of poor training estimates on model training. We find that weighting examples by confidence improves performance. We can also use the confidence measure at test time, creating an ensemble method that mediates between a spatial cue based algorithm and a model that was bootstrapped from that algorithm. This ensemble outperforms either approach by itself. 

\vfill\pagebreak
\balance

\bibliographystyle{IEEEtran_nourl}
\bibliography{refs}

\end{document}